\documentclass[a4paper,fleqn,usenatbib]{mnras}

\usepackage{mathptmx}

\usepackage[T1]{fontenc}
\usepackage{ae,aecompl}

\usepackage{graphicx}	
\usepackage{amsmath}	
\usepackage{amssymb}	

 \usepackage{times}
  \usepackage{lscape}
  \usepackage{rotating}

\newcommand{\rsun}{$R_{\odot}$}
\newcommand{\msun}{$M_{\odot}$}

\title[Doppler tomography of HD\,170582]{Doppler tomography of the Double Periodic Variable HD\,170582 at low and high stage}

\author[Mennickent, Zharikov, Cabezas et al. ]
  {R.E. Mennickent$^{1}$\thanks{E-mail: rmennick@astroudec.cl},  
     S. Zharikov$^{2}$, M. Cabezas$^{1}$,    G. Djura\v{s}evi\'c$^{3,4}$ 
\\
  $^1$Universidad de Concepci\'on, Departamento de Astronom\'{\i}a,
      Casilla 160-C, Concepci\'on, Chile\\
  $^{2}$  Observatorio Astron\'omico Nacional SPM, Instituto de Astronom\'{\i}a, UNAM, Ensenada, BC, Mexico \\
   $^{3}$ Astronomical Observatory, Volgina 7, 11060 Belgrade 38, Serbia   \\
   $^{4}$ Isaac Newton Institute of Chile, Yugoslavia Branch
     }
\date{}

\pubyear{2016}

\begin{document}
\label{firstpage}
\pagerange{\pageref{firstpage}--\pageref{lastpage}}
\maketitle

\begin{abstract} 

HD\,170582 is  an interacting binary of the Double Periodic Variable (DPV) type, showing ellipsoidal variability with  a period of 16.87  days along with a long photometric cycle of 587 days.
It was recently studied by Mennickent et al. (2015), who found a slightly evolved B-type star surrounded by a luminous accretion disc  fed by a Roche-lobe overflowing A-type giant.
Here we extend their analysis presenting new spectroscopic data and studying  the Balmer emission lines.
We find orbitally modulated double-peak H$\alpha$ 
and H$\beta$ emissions whose strength also vary in the long-term. In addition, Doppler maps of the emission lines reveal sites of enhanced line emission in the 1st and 4th velocity quadrants, the first one consistent with the position of one of the bright zones detected by the light curve analysis.  
 We find a  difference between Doppler maps at high and low stage of the long cycle; evidence that 
 the emission is optically thicker at high state in the stream-disc impact region, possibly reflecting a larger mass transfer rate.
 We compare the system parameters with a grid of synthetic binary evolutionary tracks and find the best fitting model.
The system is found  to be semi-detached, in a conservative Case-B mass transfer stage, with age 7.68  $\times$ 10$^{7}$ yr and mass transfer rate  1.6  $\times$ 10$^{-6}$ $M_{\odot} yr^{-1}$. 
For 5 well-studied DPVs, the disc luminosity scales with the primary mass and is much larger than the theoretical accretion luminosity.

\end{abstract}




\section{Introduction}

The interacting binary HD\,170582 (BD-14 5085, ASAS ID 183048-1447.5, $\alpha_{2000}$ = 18:30:47.5, $\delta_{2000}$ = -14:47:27.8, $V$ = 9.66 mag, $B-V$ = 0.41 mag, spectral type A9V)\footnote{http://simbad.u-strasbg.fr/simbad/}  is a member of the class of Double Periodic Variables, Algol-related binaries showing a long photometric cycle lasting about 33 times the orbital period  with a period ratio mostly  between 25 and 45 (Mennickent et al. 2003, Mennickent et al. 2008, Poleski et al. 2010, Mennickent et al. 2012a, 2012b, Mennickent 2013, Garrido et al. 2013, Barr\'{\i}a et al. 2013, 2014, Pawlak et al. 2013).  Recently, Mennickent et al. (2016) showed that DPVs are those semi-detached Algols with B-type primaries of radius slightly larger than the critical radius\footnote{The critical radius just allows the stream coming from the secondary  to hit the primary. A smaller primary radius is traditionally assumed for the existence of an accretion disc.} and also that they host stable accretion discs \citep{2015arXiv151005628M}.

In a recent study Mennickent et al. (2015, hereafter M15)  disentangled the light curve of HD\,170582 into an orbital part, determining a period of 16.87 days and revealing orbital ellipsoidal variability with unequal maxima, and a long cycle of 587 days, showing quasi-sinusoidal changes with amplitude $\Delta V$= 0.1 mag. They modeled the orbital light curve with a cool evolved star of $M_{2}$ = 1.9 $\pm$ 0.1 $M_{\odot}$, $T_{2}$ = 8000 $\pm$ 100 $K$ and $R_{2}$ = 15.6  $\pm$ 0.2 $R_{\odot}$, and an early B-type dwarf of $M_{1}$ =  9.0 $\pm$ 0.2 $M_{\odot}$
 surrounded by a geometrically and optically thick accretion disc of radial extension  
20.8 $\pm$ 0.3 \rsun,  contributing about 35\% to the system luminosity  in the $V$-band. M15 fit the light curve asymmetries with two extended regions located at opposite sides of the disc rim, and hotter than the disc by 67\% and 46\%. One of these regions can be associated to the  impact with, and penetration of the gas stream into the accretion disc.  

The origin of the long cycle in DPVs still is not understood, but it could be related to episodic mass loss from the system driven possibly by a bipolar wind (Mennickent et al. 2008, 2012b). This result  shows the importance of imaging the circumstellar material 
with the aid of tomographic reconstructions, which are still scarce for the class of DPVs   (Atwood-Stone et al. 2012, Richards et al. 2014, Mennickent et al. 2012b).


In this paper we complement the work by M15, studying the long-term variability of the Balmer emission. We also get insights on the line emissivity distribution at high and low stage with the aid of the technique of Doppler tomography. 
We also investigate for the first time the evolutionary stage of this binary determining its age and  mass transfer rate and provide insights on the luminosity of the discs surrounding the primaries of DPVs.
We divide the paper as follows: in Section 2 we present the spectroscopic data used in this paper, in Section 3 we give insights on emission line profile variability. In the same section we build Doppler tomograms for  the H$\alpha$ and  H$\beta$ emission lines, discuss the system evolutionary stage, obtain a value for the mass transfer rate, and present a study of the luminosity of the DPV accretion discs. We give a discussion in Section 4 and our conclusions are presented in Section 5.

\section{Spectroscopic data}

Most of the observing material considered in this paper was described by M15, it consists 
 of 136 echelle optical spectra obtained between years 2008 and 2013 with resolving power $R \sim$ 40,000,  at the observatories Las Campanas with the echelle spectrograph  (DuPont telescope), 
Cerro Tololo with the CHIRON spectrograph (1.5m telescope)  and La Silla with the CORALIE spectrograph (EULER telescope). 
In addition to the aforementioned observing material, new spectra are studied in this paper.

The  new spectroscopic data were obtained using the Echelle Spectrograph \citep{Levine1995} attached to the 2.12 m telescope of
the Observatorio Astron\'omico Nacional at San Pedro M\'artir (OAN SPM), Mexico. This spectrograph provides spectra spread over 27 orders, covering the spectral range of  3500-7105~\AA\, 
with  a spectral resolution power of $R$ = 18000. A  ThAr lamp was used for wavelength calibration. The spectra were reduced by a standard way using the echelle package in IRAF\footnote{IRAF is distributed by the National Optical Astronomy Observatories,
 which are operated by the Association of Universities for Research
 in Astronomy, Inc., under cooperative agreement with the National
 Science Foundation.}, 
 including flat and bias correction, wavelength calibration and order merging. All spectra discussed in this paper are  normalized to the continuum and 
the RVs are heliocentric ones.  We use the orbital and long-cycle ephemerides for the binary provided by M15.
 The log of observation and characteristics  of the new spectra  are given  in Table\,1. 


\begin{table}
\centering
 \caption{Summary of new spectroscopic observations obtained with the echelle spectrograph at the 2.1m telescope of  the San Pedro M\'artir Observatory. At each epoch one exposure of 1200 s was secured. The heliocentric julian day (HJD) at mid-exposure  is given. The $S/N$ ratio is calculated in the continuum around H$\alpha$. $\Phi_{\rm{o}}$ and $\Phi_{\rm{l}}$ refer to the orbital and long-cycle phase, respectively, and are calculated according to the ephemerides given in M15.}
 \begin{tabular}{@{}lccccc@{}}
 \hline
UT-date &UT-middle &S/N  &HJD &$\Phi_{\rm{o}}$& $\Phi_{\rm{l}}$\\
\hline
2014-09-19 & 04:17:24 &60 & 2456919.67296 &0.8353 &0.2603\\
2014-09-19 &04:37:36 &68 & 2456919.68698 &0.8361 &0.2603\\
2014-09-19 &04:57:48 &62 & 2456919.70101 &0.8369 &0.2603\\
 2014-11-02&02:02:16 &79 &2456963.57498 &0.4373 &0.3351\\
 2014-11-02&02:22:33 &82 &2456963.58906 &0.4381 &0.3351\\
 2014-11-02&02:49:57&92 &2456964.60801 &0.4985 &0.3368\\
\hline
\end{tabular}
\end{table}

\section{Results}

\subsection{Spectroscopy}

\subsubsection{Spectral disentangling and donor-subtracted spectra}

The following light sources are present in HD\,170582: the donor, the gainer, the disc and possibly the gas stream. 
In order to remove the donor from the spectra we assume that its contribution to the total light adds to the other  sources and it is characterized by a synthetic spectrum 
with $T_2$ = 8000 $K$,  log\,$g_2$ = 1.7, projected rotational velocity 44 km s$^{-1}$ and microturbulent velocity  1 km s$^{-1}$.  This spectrum was selected by M15 from a grid of spectra calculated using the SYNTHE code plus atmospheric models computed with the line-blanketed local thermodynamic equilibrium ATLAS9 code (Kurucz 1993).

 
The donor-subtracted spectra were constructed removing the donor synthetic spectrum from the observed spectra.  Firstly, the synthetic spectrum was Doppler corrected and scaled according to the  contribution of the secondary star at a given orbital phase and at the given wavelength range. For normalized fluxes at given orbital and long-cycle phases ($\Phi_{o}$ and 
$\Phi_{l}$,   respectively):

\begin{equation}
f_{res} ( \lambda, \Phi_o, \Phi_l)  = f (\lambda, \Phi_o,  \Phi_l) - p(\Phi_o, \lambda_c) \times f_d (\lambda, \Phi_o)
\end{equation}

\noindent
where $f_{res}$ is the donor-subtracted flux, $f$ the observed flux, f$_{d}$ the  synthetic  spectrum of the donor, $p$ is its 
fractional contribution  according to our model and $\lambda_{c}$ the representative wavelength where this factor is calculated. Theoretical $p$-factors were computed from the M15 model; they correctly account for the variable  projected area of the donor and its larger flux contribution at longer wavelengths (Fig.\,1). This method effectively removed the main contribution of the donor from the observed spectra. The result was a set of  disentangled spectra that were normalized to the new continuum. 
 
 \begin{figure}
\scalebox{1}[1]{\includegraphics[angle=0,width=8cm]{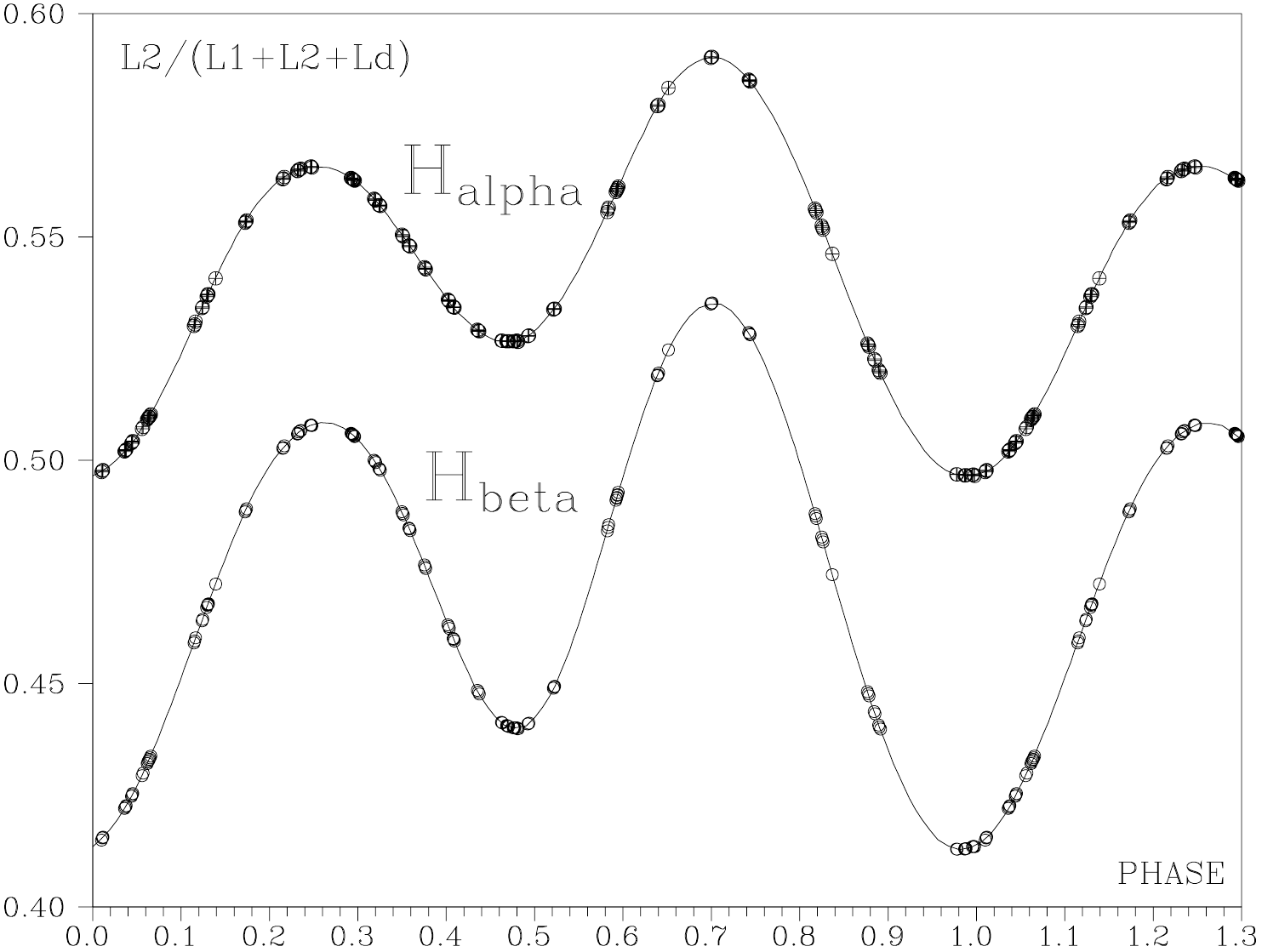}}
\caption{Fractional donor flux contribution  at H$\alpha$ and H$\beta$ for the orbital phases covered by the observations. $L_1$, $L_2$ and $L_d$ refer to the gainer, donor and disc fluxes, respectively.
}
  \label{x}
\end{figure}

\begin{figure}
\scalebox{1}[1]{\includegraphics[angle=-90,width=8cm]{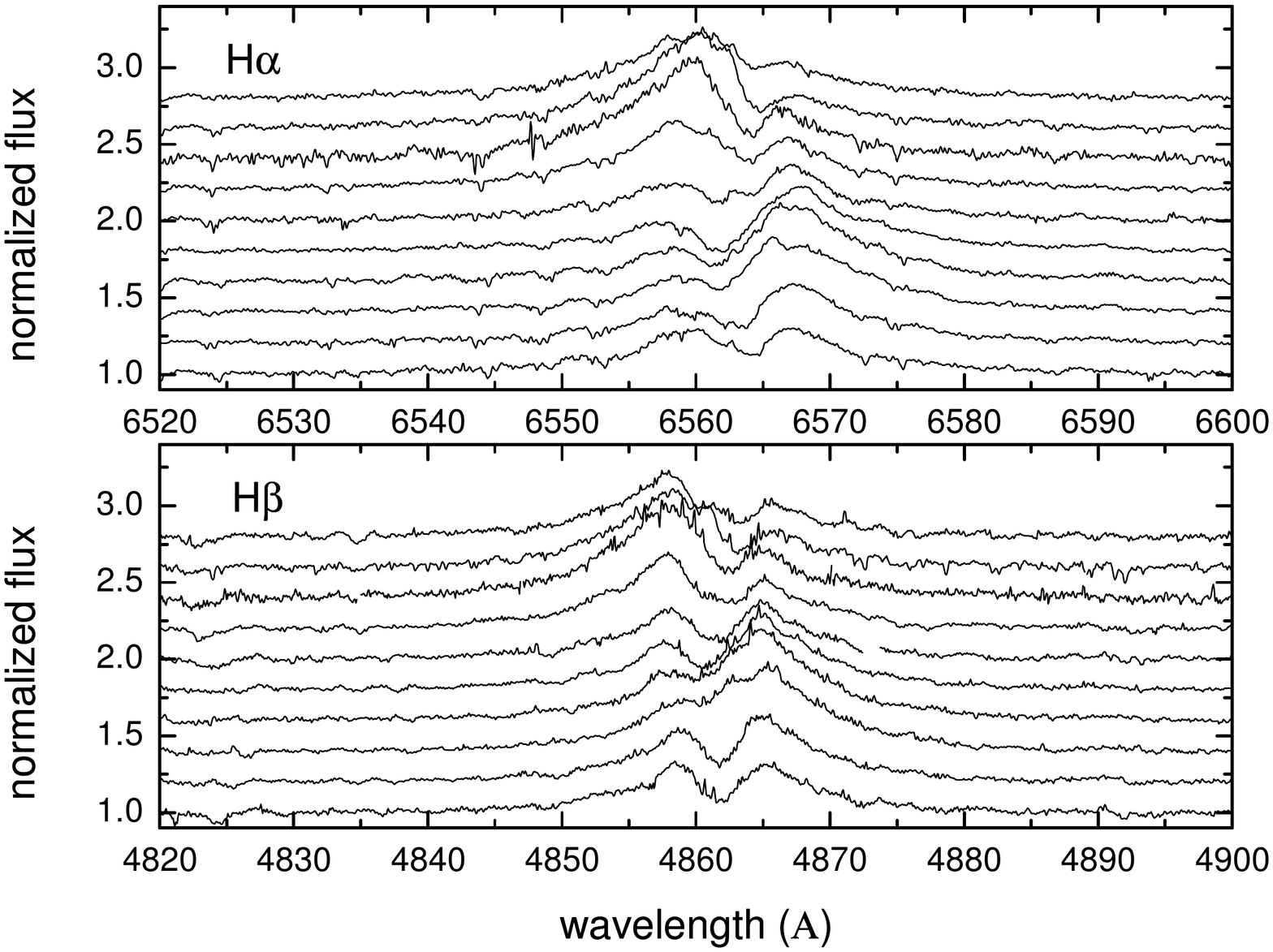}}
\caption{Line profile variability.  From bottom to top we show averaged CHIRON profiles at orbital phases 0.00, 0.10, 0.20, .... up to 0.90. 
}
  \label{x}
\end{figure}

\begin{figure}
\scalebox{1}[1]{\includegraphics[angle=0,width=8cm]{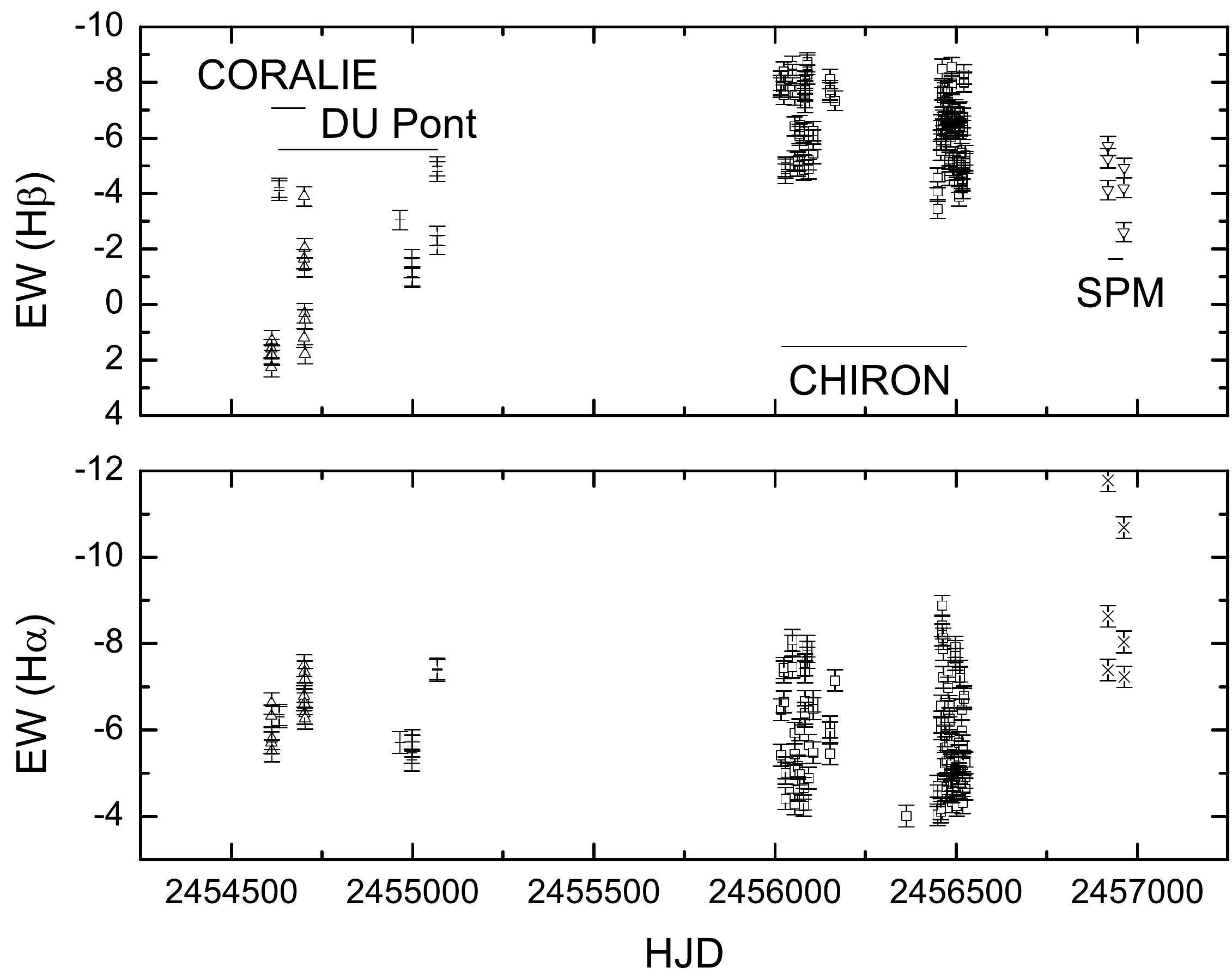}}
\caption{H$\alpha$ and H$\beta$ equivalent widths. Epochs covered with different instruments are indicated.}
  \label{x}
\end{figure}

\begin{figure}
\scalebox{1}[1]{\includegraphics[angle=0,width=8cm]{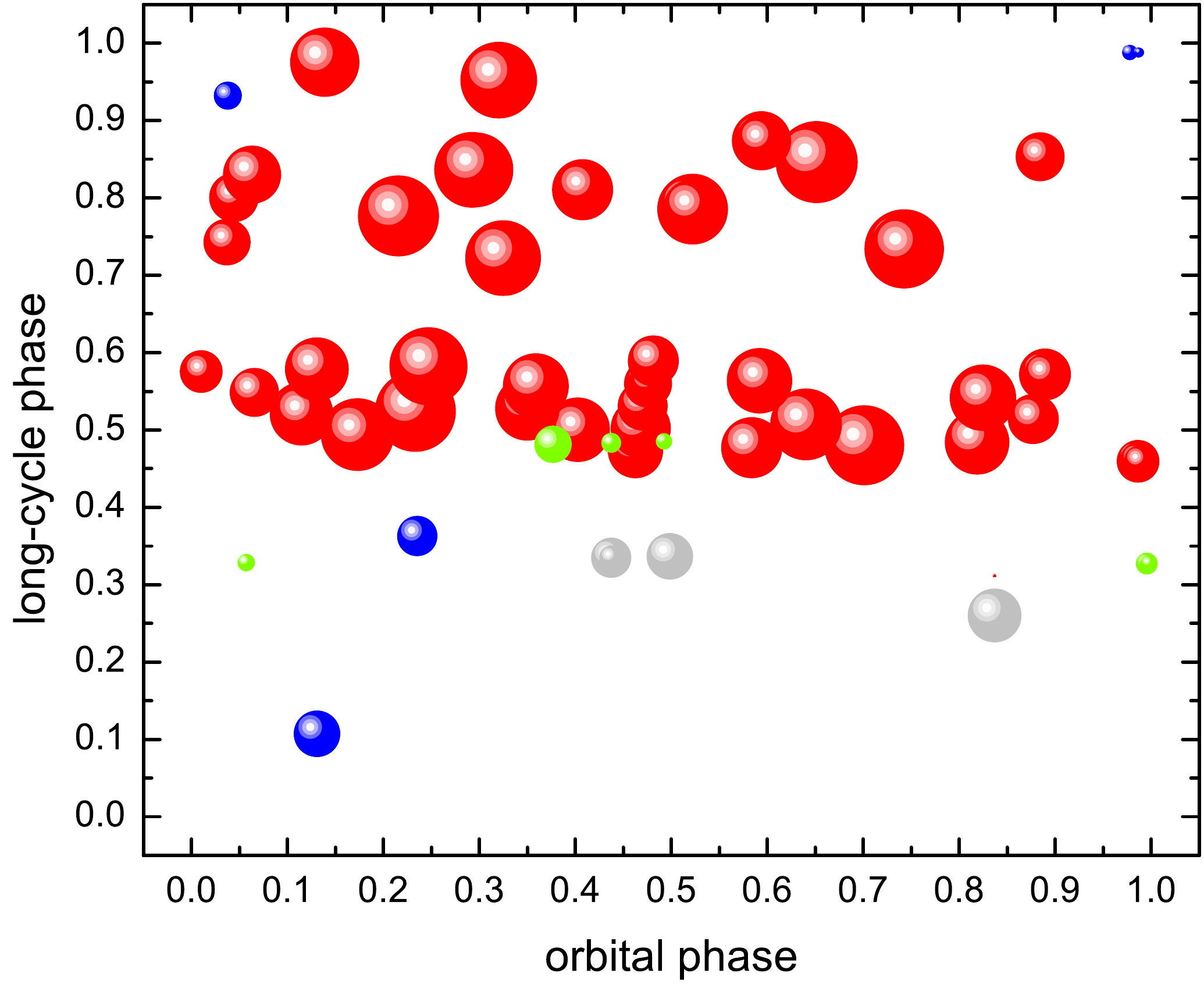}}
\scalebox{1}[1]{\includegraphics[angle=0,width=8cm]{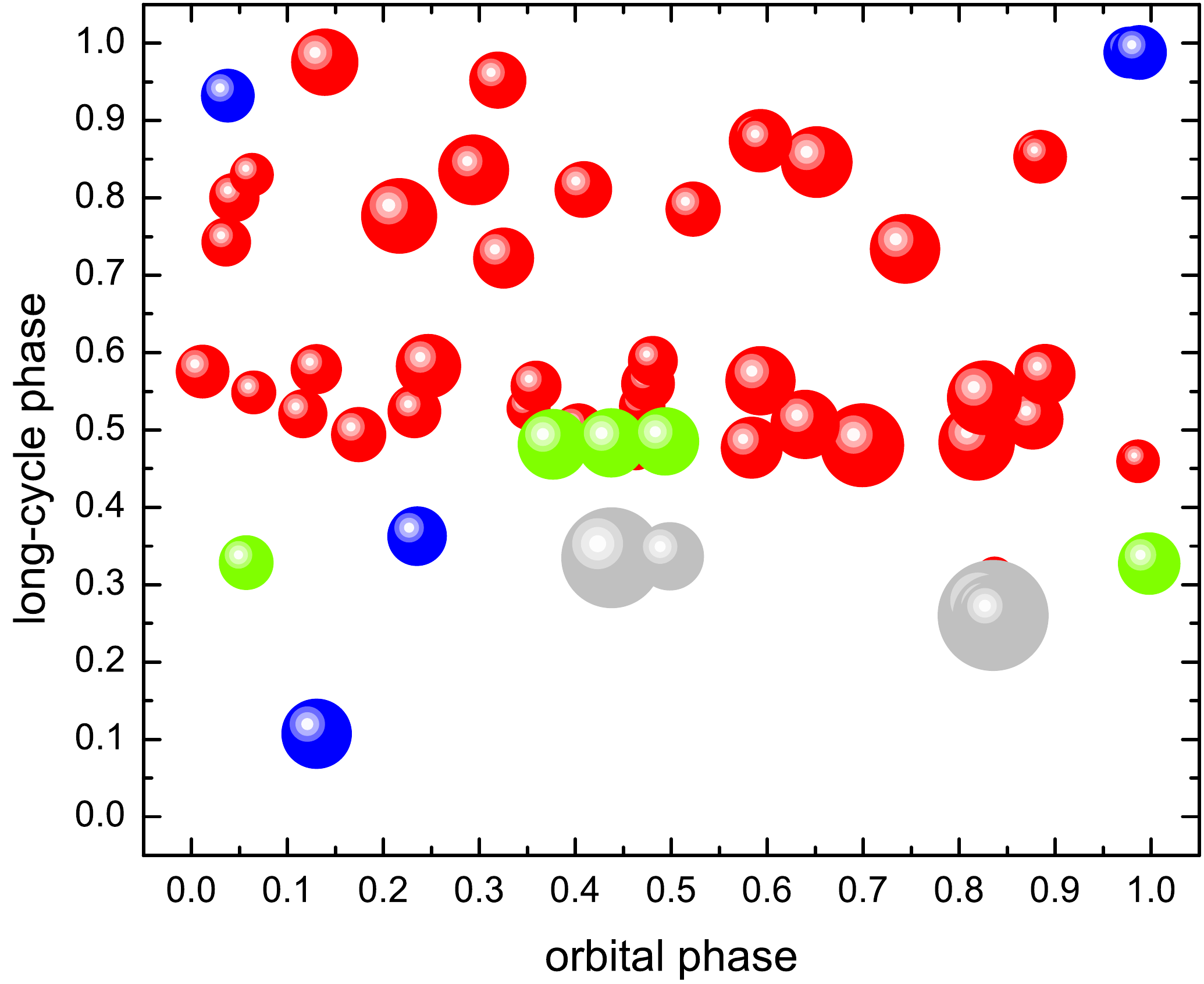}}
\caption{Distribution of spectra in the orbital-phase versus long-cycle phase diagram. Chiron (red), SPM (gray), CORALIE (green) and DuPont (blue) data are indicated.  
The symbol size scales with the absolute value of H$\beta$ (top) and H$\alpha$ (bottom) equivalent widths.
}
  \label{x}
\end{figure}
   

Disentangling of the gainer and the disc was not intended since there is no model for the disc absorption lines,  and there is evidence that its spectrum combine in a complex way with the stellar spectrum of the primary (M15). We find in the residual spectra H$\alpha$ and H$\beta$ double emission lines and a central absorption moving across the emission regularly during the whole cycle (Fig.\,2).

\begin{figure}
\scalebox{1}[1]{\includegraphics[angle=0,width=8cm]{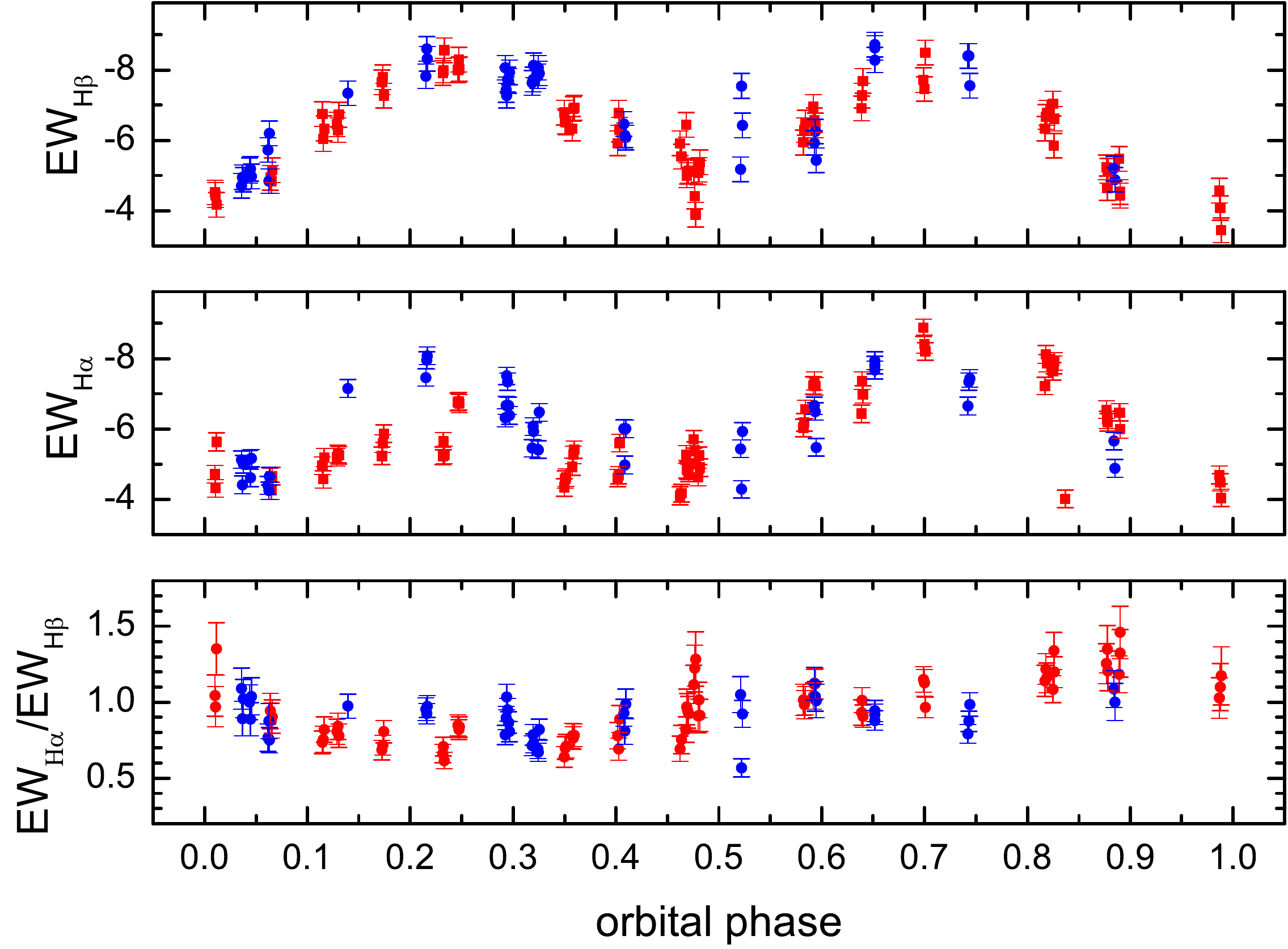}}
\caption{Orbital variability of
emission line equivalent widths  for CHIRON data.  Data for $0.3$ $<$ $\Phi_{l}$ $\leq$ 0.6 (red) and  $0.6$ $<$ $\Phi_{l}$ $\leq$ 1.0 (blue) are separately shown. }
  \label{x}  
\end{figure}

\subsubsection{Long-term spectroscopy}

The long-term behavior of the Balmer emission, as measured by the equivalent width,  is shown in Fig.\,3. While H$\alpha$  practically does not show large long-term variability  (except for two spectra taken in 2014), H$\beta$ 
emission is stronger during 2012-2013 (CHIRON observations) and weaker during 2014 (SPM data) and 2008-2009 (CORALIE \& DuPont observations). 

The investigation of the long-cycle variability is limited by timings of our dataset, which are scarce during  the first half of the long cycle; although most SPM/Du-Pont/CORALIE data are found in this range they consist  of  very few epochs.
In general, the H$\beta$ emission is lower during the first half of the long-cycle, but H$\alpha$ does not show the same pattern (Fig.\,4).

For CHIRON data the Balmer emission shows a double-wave modulation during the orbital cycle, closely following the light curve described by M15; 
 it is maximum at quadratures, i.e. when continuum emission is maximum (Fig.\,5). The ratio between the equivalent widths of H$\alpha$ and H$\beta$ describes a modulation with the orbital cycle with a maximum around $\Phi_{o}$ = 0.9. Interestingly, there is no evidence for changes in the emission line strength with the long-cycle during the interval covered by CHIRON data (roughly $0.45$ $<$ $\Phi_{l}$ $<$ 1.0).

\subsection{Emission line orbital variability}

We measured the position of the blue and red peak, the central absorption (CA)
and  the intensities of the H$\alpha$ and H$\beta$ lines. This was made by directly measuring the CA positions with the cursor  and fitting each peak with a gaussian using the IRAF "splot" task. The method of deblending the whole profile with a double-gaussian function produced spurious results, specially for the CA, and was not considered. The measurements show that the central absorption moves quasi-sinusoidally in both lines with total amplitude about 180 km s$^{-1}$ in H$\alpha$ and 210 km s$^{-1}$ in H$\beta$.  The peaks follow a more complex pattern (Fig.\,6).
The CA reaches maximum receding velocity around $\Phi_{o}$ = 0.85 and maximum approaching velocity around $\Phi_{o}$ = 0.30.
The H$\alpha$ peak separation is 390 $\pm$ 60 (std) km s$^{-1}$ and for H$\beta$  is 330 $\pm$ 30 km s$^{-1}$.
For H$\alpha$ the full width at zero intensity is 2555 $\pm$ 260 km s$^{-1}$ and for  H$\beta$ is 2660 $\pm$ 370 km s$^{-1}$. 
The peaks intensity varies in anti phase during the cycle  (Fig.\,7)
and the central absorption scatters around a level about 20\% below the continuum.  The  red peak is larger around $\Phi_{o}$ = 0.25 and minimum around $\Phi_{o}$ = 0.80, while the opposite happens for the blue peak (Fig.\,7).


\subsection{Doppler tomography}

 We use the Doppler tomography  technique  to  probe 
the structure of the accretion disc and circumstellar material in HD\,170582. 
 Recently, 
this method has been successfully applied to Algols revealing their importance to 
obtain images of gravitational and magnetic phenomena in these interacting binaries \citep{2014ApJ...795..160R}. 
%
%
The general description of the method can be found in  the original paper of  \citet{1988MNRAS.235..269M}. Briefly, The  Doppler tomography technique uses the information encoded in spectral line profiles taken at different orbital phases to calculate a distribution of emission over the binary. Doppler tomography provides a quantitative mapping of optically thin line forming regions  in velocity space. 


The Doppler maps for H$\alpha$ and H$\beta$ emission lines 
were constructed 
using the maximum entropy method \citep{1994A&A...289..983L} as implemented by \citet{1998astro.ph..6141S}\footnote{The code and its description are available at http://www.mpa-garching.mpg.de/$\sim$henk/pub/dopmap/}.
Our time-resolved spectral observations allowed us to cover practically uniformly  all orbital phases of the system within an average of $\sim7\pm3$ spectra  per 0.05 phase step   of the orbital period  (see Fig.\,4). 
%
The system parameters of
$M_{1}$ = 9.0 M$_{\odot}$, mass ratio $q$ = 0.21, inclination angle 67\fdg4 and orbital period 16\fd87 from M15  are used to 
plot the positions of the stellar components, the mass transfer stream and  its Keplerian velocity 
and the outer  edge  of the accretion  disc. 
We also separately considered  spectra taken at the high stage (0.7 $\leq$ $\Phi_l$ $<$ 1.2) and the low stage (0.2 $\leq$ $\Phi_l$ $<$ 0.7)\footnote{The high state of the system corresponds to the system brightest increasing along of the long period, the low state is the opposite case, see Fig.\,1 from M15}. In  practice, due to data sampling, most spectra are in the ranges 0.7 $<$ $\Phi_l$ $<$ 1.0 and 0.4 $<$ $\Phi_l$ $<$ 0.6 but with a very good coverage of the orbital cycle (see Fig.\,4). The resulting Doppler maps are presented in Figs.\,8 and 9.


We find that the Balmer line emission is not formed in a disc, but in a horseshoe structure located in the upper velocity  hemisphere. In addition, bright structures are detected in the first (upper left) quadrant and also in an elongated region in the fourth (upper right) quadrant. Practically no emission is found at the second and third quadrants. 
 Two bright regions are found at angular distances  $\sim$ 320\degr and $\sim$ 10\degr $-$ 35 \degr respectively, measured from the line joining the stellar centers as measured in direction opposite to the orbital motion. The first one can be  associated with the bright spot  
detected in the LC analysis by M15 at $\sim$ 334\degr, but the second large bright region is different from the bright spot detected in the continuum light at $\sim$ 135\degr (M15). The emissivity (compared to the local continuum) is larger in H$\alpha$ 
than in H$\beta$ and follows a similar horseshoe pattern. On the other hand, the emission ratio H$\alpha$/H$\beta$ is not  axisymmetrical, which reveals  spatially variable physical conditions in the circumstellar material. 

Particle-dynamic simulations performed for  close binary systems  allow us to  interpret the bright zones detected in the Doppler maps. \citet{1998MNRAS.300...39B}  found that the gas stream penetrates the disc forming a hotline  that could be related to the bright region detected in the 1st quadrant. In addition, \citet{2001PThPh.106..729F}   shows a shock region in the 4th quadrant produced by the impact of disc material onto this hotline. 

 There is practically no differences among the H$\beta$ Doppler maps obtained at high and low stage  and almost the same happens for H$\alpha$, except in the 1st quadrant, where emission is  notably lower during high state. This change in the ratio of H$\alpha$ to H$\beta$ emission occurs in the stream-disc impact region, and  might indicate a change in the physical characteristics of the material, from an optically thin (at the low stage) to an optically thick (at high stage) condition.\footnote{The Balmer emission decrement is usually taken as an indicator of optical thickness in gas excited by photoionization and electronic recombination}.

 Since the present analysis  is based on irregularly sampled spectra obtained over a relatively long time-span, some possible intrinsic variability might not be resolved by the Doppler maps, being masked during the process of phase-binning the spectra.
The current volume of the data does not allows us to study this possible variability with Doppler tomography.

\begin{table}
\centering
 \caption{The stellar parameters used in this paper   to find the best evolutionary model for HD\,170582. Data are from Mennickent et al. (2015). } 
 \begin{tabular}{@{}lclc@{}} \\
 \hline
 \hline
{\rm quantity}  &{\rm value} &  {\rm quantity} &{\rm value} \\
 \hline
$M_{c}$  &1.9 $\pm$ 0.1 M$_{\odot}$& $M_{h}$  &9.0 $\pm$ 0.2 M$_{\odot}$ \\
$T_{c}$ & 8000 $\pm$ 100 $K$& $T_{h}$ & 18000 $\pm$ 1500 $K$\\
log $T_{c}$ &3.903 $\pm$ 0.005   &log $T_{h}$ &4.255 $\pm$ 0.037    \\
$L_{c}$ & 863 $\pm$ 80 L$_{\odot}$ &$L_{h}$ & 2858 $\pm$ 529 L$_{\odot}$ \\
log $L_{c}/ L_{\odot}$ &2.936 $\pm$ 0.040 &log $L_{h}/ L_{\odot}$ &3.456 $\pm$ 0.080\\
$R_{c}$ &15.6 $\pm$ 0.2  R$_{\odot}$ &$R_{h}$ &5.5 $\pm$ 0.2   R$_{\odot}$  \\
\hline
\end{tabular}
\end{table}

\begin{table}
\centering
 \caption{The parameters of the Van Rensbergen et 
al.  (2008) model that best fit the HD\,170582 data of  Table 2. The 
hydrogen and helium core mass fractions are given for the cool  ($X_{cc}$ and $Y_{cc}$)  and hot   ($X_{ch}$ and $Y_{ch}$) star. The errors represent 
the grid step at a given parameter.} 
 \begin{tabular}{@{}lclc@{}} \\
 \hline
 \hline
{\rm quantity}  &{\rm best model} &{\rm quantity}   & {\rm best model} \\
 \hline
age & 7.6778(5)E7   yr    &       period &16.89 $\pm$ 0.17 d  \\ 
$M_{c}$  &1.922 $\pm$ 0.007 M$_{\odot}$   &$M_{h}$  &9.474 $\pm$ 0.008 M$_{\odot}$ \\
$dM_{c}/dt$ &-1.6(2)E-6   M$_{\odot}$ yr$^{-1}$ &  $dM_{h}/dt$ &1.6(2)E-6 M$_{\odot}$ yr$^{-1}$ \\
$T_{c}$ &6918 $\pm$ 16 $K$ &$T_{h}$ & 23120 $\pm$ 0 $K$\\
log $T_{c}$ &3.840 $\pm$ 0.001    &log $T_{h}$ &4.364 $\pm$ 0.000   \\
$L_{c}$ &507 $\pm$ 12 L$_{\odot}$ &$L_{h}$ &7656$^{+8}_{-0}$   L$_{\odot}$ \\
log $L_{c}/L_{\odot}$ &2.705 $\pm$ 0.010  &log $L_{h}/L_{\odot}$ &3.8840 $\pm$ 0.0005 \\
$R_{c}$ &15.65 $\pm$ 0.08  R$_{\odot}$  &$R_{h}$ &5.452 $\pm$ 0.002   R$_{\odot}$\\
$X_{cc}$ &0.000 $\pm$ 0.000  &$X_{ch}$ &0.494 $\pm$ 0.000\\
$Y_{cc}$ & 0.980 $\pm$ 0.000 & $Y_{ch}$ &0.486 $\pm$ 0.000  \\
\hline
\end{tabular}
\end{table}

\begin{figure}
\scalebox{1}[1]{\includegraphics[angle=0,width=8cm]{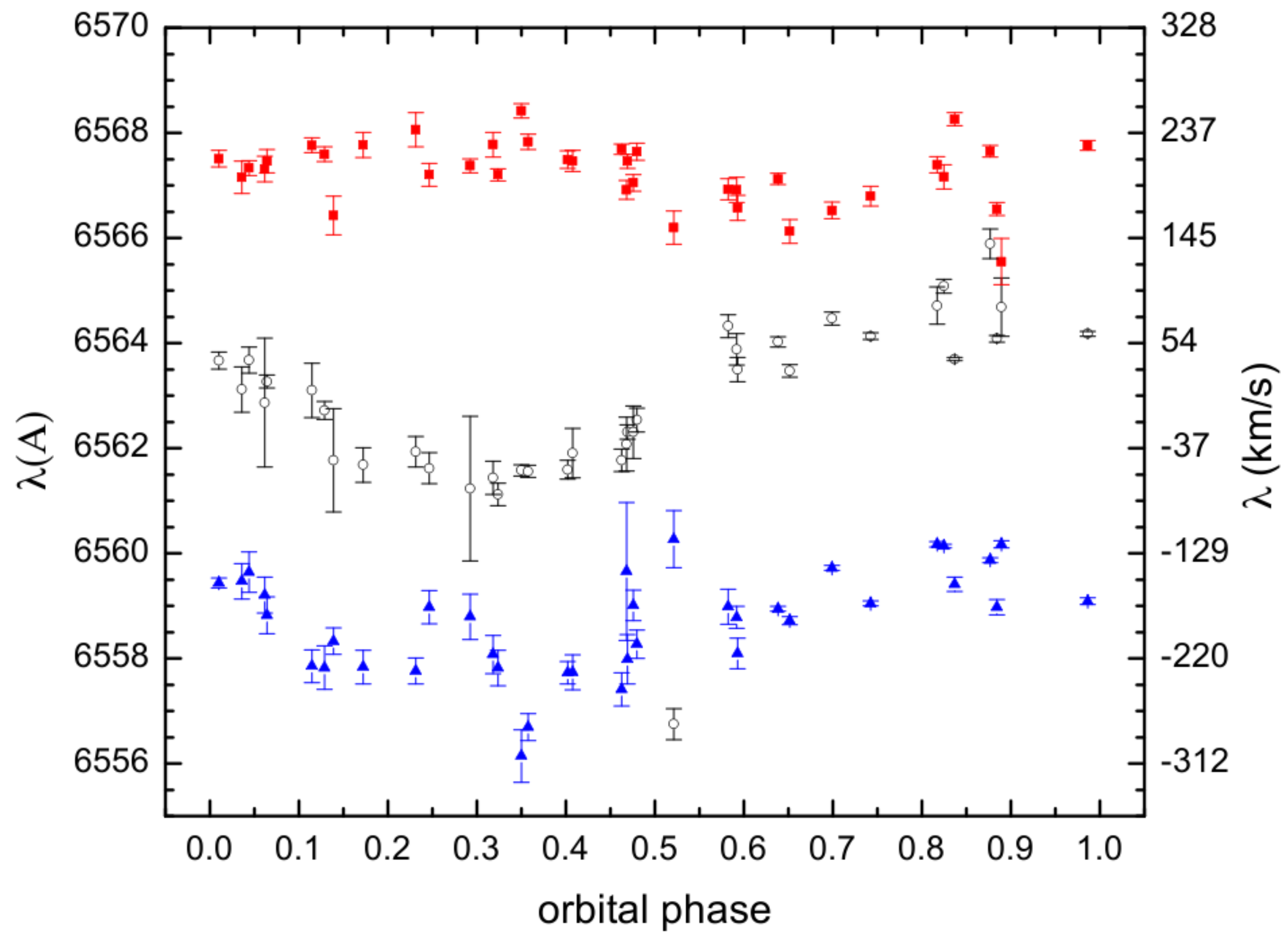}}
\scalebox{1}[1]{\includegraphics[angle=0,width=8cm]{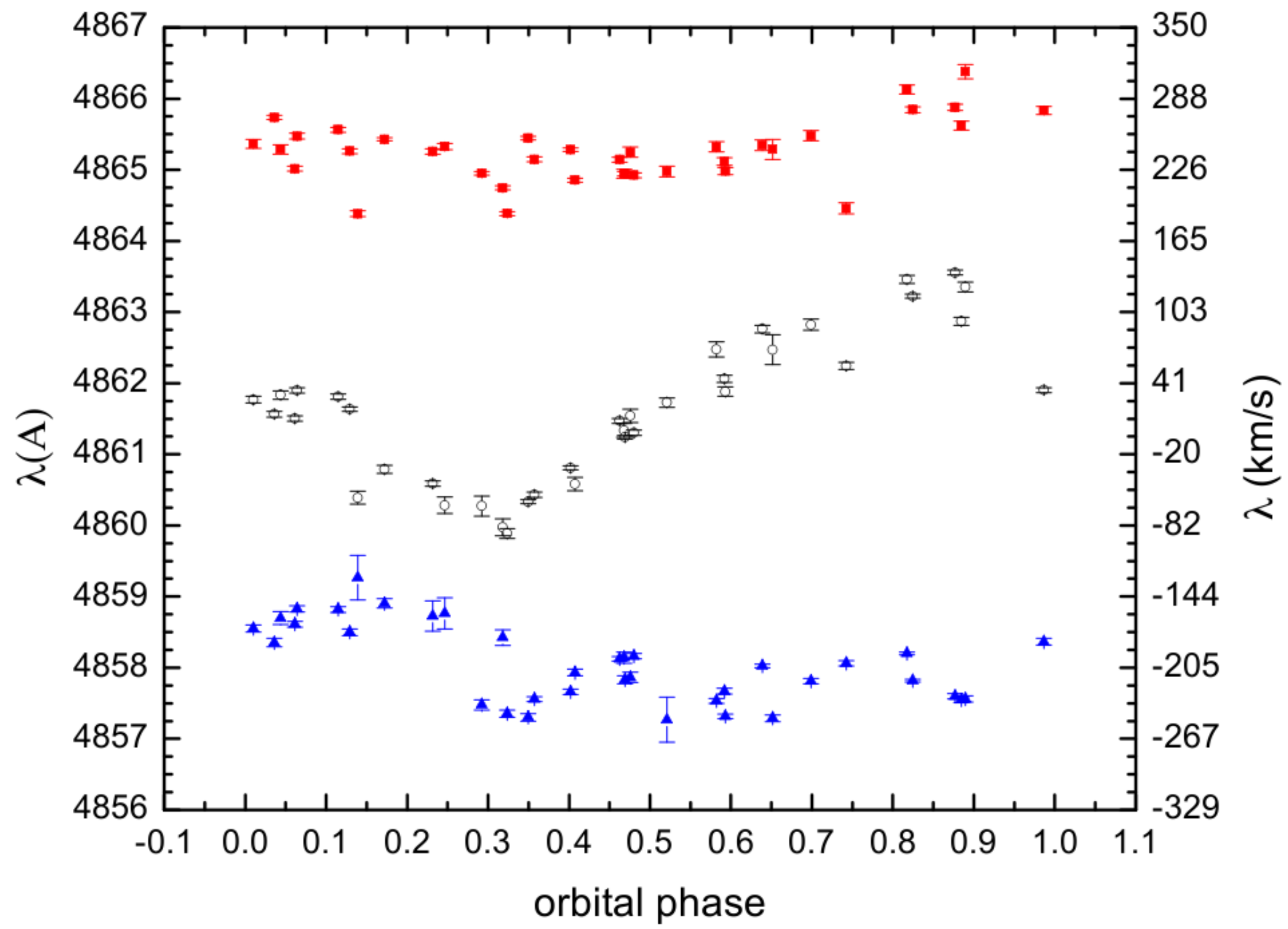}}
\caption{Positions of the peaks and central absorption for the H$\alpha$ (top) and H$\beta$ (bottom) emission lines. The velocity scale refers to laboratory rest wavelengths. 
}
  \label{x}
\end{figure}

\begin{figure}
\scalebox{1}[1]{\includegraphics[angle=0,width=8cm]{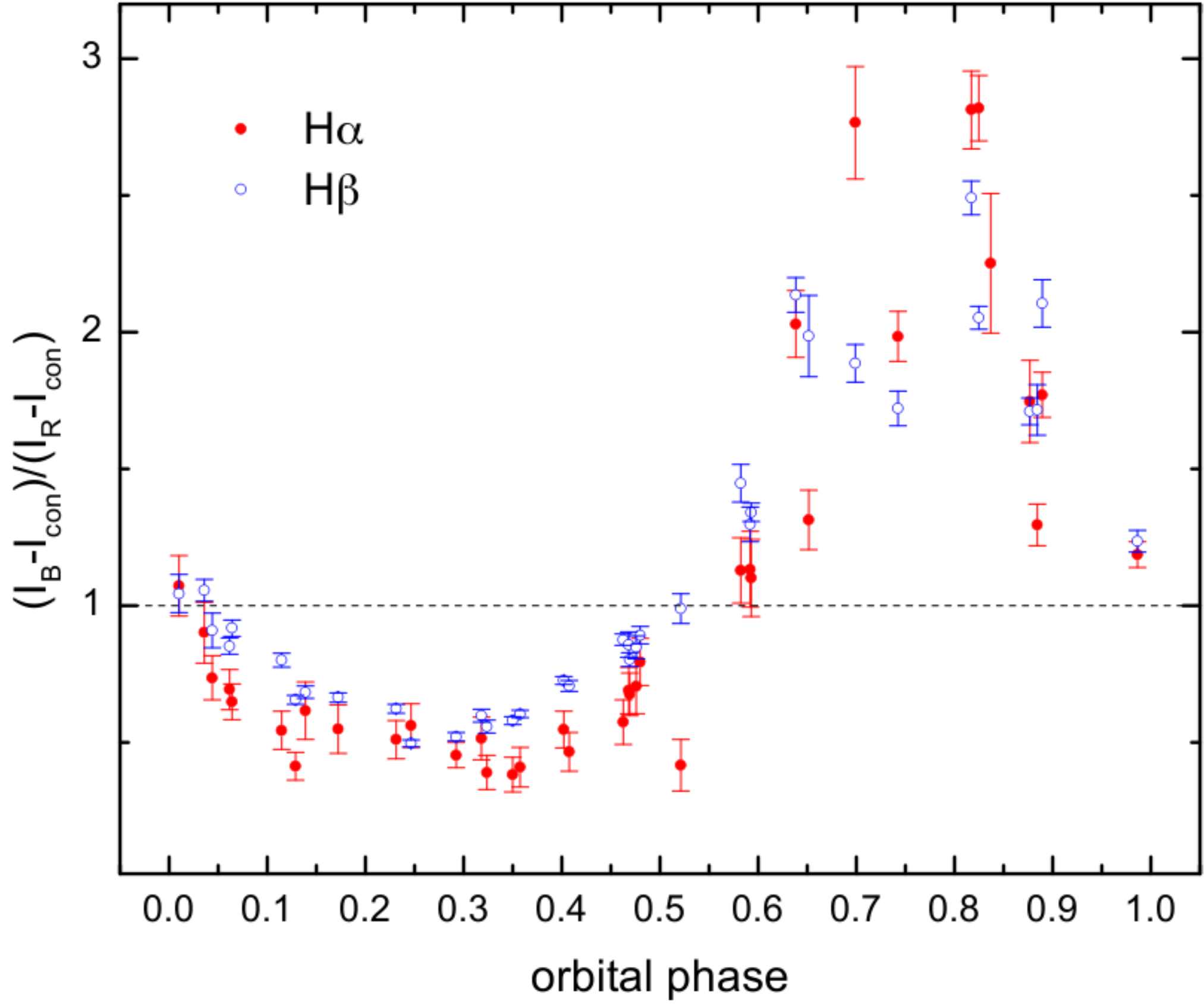}}
\caption{Ratio between the blue and red emission peak intensity referred to the continuum level.
}
  \label{x}
\end{figure}

\subsection{Evolutionary stage, age and mass transfer rate}

In this Section we compare the system parameters with those predicted by binary evolution models including epochs of non-conservative evolution following the method described by Mennickent (2014).
The stellar and system parameters given by M15 were chosen as representative for the system, and are given in Table\,2; in this paper we use subscript h and c for the hot and cool star, respectively. 
In our study the orbital period of 16.8722(17) days is also taken from M15. We discuss the effect of the uncertainties associated to the system parameters in our methodology. 

We inspected the 561 evolutionary tracks  by Van Rensbergen et 
al.  (2008), available at the Center de Donn\'ees Stellaires (CDS), searching for the best match for the system parameters.
Models with strong and weak tidal interaction were studied, along with conservative and non-conservative models.
Following Mennickent et al. (2012a), a  multi-parametric fit was made with the  synthetic ($S_{i,j,k}$)  and observed ($O_{k}$) parameters  stellar mass, temperature, luminosity and radii, and the system orbital period,  where $i$ (from 1 to 561) indicates the synthetic  model, $j$ the time $t_{j}$ and $k$  (from 1 to 9)  the  stellar or orbital parameter.
 Non-adjusted parameters were mass loss rate, Roche lobe radii, chemical composition, age and fraction of accreted mass lost by the system.
 For every synthetic model $i$ we calculated the quantity $\chi^{2}_{i,j}$ at every $t_{j}$ defined by

\begin{equation}
\chi^{2}_{i,j} \equiv (1/N) \Sigma_{k} w_{k}[(S_{i,j,k}-O_{k})/O_{k}]^{2},
\end{equation}
\noindent
where $N$ is the number of observations (9) and $w_{k}$ the statistical weight of the parameter $O_{k}$, calculated as

\begin{equation}
w_{k} = \sqrt{O_{k}/\epsilon(O_{k})},
\end{equation}
\noindent
where $\epsilon(O_{k})$ is the error associated to the observable $O_{k}$.
The model with the minimum $\chi^{2}$ corresponds to the model with the best evolutionary history of HD\,170582. The absolute minimum $\chi^{2}_{min}$  gives the age of the system along with the theoretical stellar and orbital parameters. The high accuracy of the orbital period dominates the search for the best solution in a single evolutionary track, but the others parameters play a role when comparing tracks corresponding to different initial stellar masses and orbital periods.

We find the best model for the system with a $\chi^{2}$ minimum at 0.020. Parameters for this solution are given in Table\,3. 
The absolute $\chi^{2}$ minimum corresponds to the conservative 
model with initial masses of 6  M$_{\odot}$ and 5.4 M$_{\odot}$ and initial orbital period of 3.0  days.
The best model gives a temperature of 23 kK and a mass of 9.47 M$_{\odot}$ for the B star, which turns  out to be slightly evolved with  a core consisting of hydrogen and helium in almost equal fractions.
The best fit also indicates that  HD\,170582  is found inside a burst of mass transfer, the second one in the lifetime of this binary (Fig.\,11). The donor is an inflated ($R_{c}$ = 15.6 R$_{\odot}$) and evolved 1.92 M$_{\odot}$ star with its core completely exhausted of hydrogen. According to the best model the system has now an age of 7.68 $\times$ 10$^{7}$ yr,  a mass transfer rate $dM_{c}/dt$ =  -1.6 $\times$ 10$^{-6}$  M$_{\odot}$ yr$^{-1}$ and a mass ratio $q$ = 0.20. The orbital period rate of change, according to the best model,  is about 2.9 seconds per year, impossible to detect with the accuracy of the derived photometric period (146.9 seconds, M15).

Deviations of the best fit parameters from those obtained from the LC model are $\Delta M_h$ = 0.47 M$_{\odot}$, $\Delta M_c$ = 0.02 M$_{\odot}$, $\Delta R_h$ = -0.05 R$_{\odot}$,   $\Delta R_c$ = 0.05 R$_{\odot}$, $\Delta T_h$ = 5120 K and  $\Delta T_c$ = 1082 K, i.e. up to 1\% for the radii,  5\% for the masses and 20\% for the temperatures.


Although formal errors can be obtained for the derived parameters, especially age and  $dM/dt$, from the above described methodology, we notice that the limited grid of models and ad-hoc assumption of mass loss in the non-conservative cases are intrinsic  limitations.
To illustrate possible uncertainties we notice that among the 10 better models, i.e. those with lower $\chi^{2}$ (between 0.020 to 0.029), the range in age is 7.3 $\times$ 10$^{7}$ yr to 7.7 $\times$ 10$^{7}$ yr and in $dM_{c}/dt$ is -2.2 $\times$ 10$^{-5}$  M$_{\odot}$ yr$^{-1}$  to -1.1 $\times$ 10$^{-6}$  M$_{\odot}$ yr$^{-1}$.

\subsection{On the  luminosity of DPV discs}

It is possible that discs of DPVs are different from classical accretion discs since the primary star, accelerated by the tangential impact of the gas stream \citep{2015arXiv151005628M}, might be rotating at  the  critical velocity, impeding further accretion. In addition, DPV discs are smaller than the tidal radius \citep{2015arXiv151005628M}, something that could indicate young discs or efficient mass loss from the system. In any case, few  studies have been performed of these discs and in this Section we explore their properties, especially concerning their luminosity.

The luminosity of an accretion disc is given by:
\begin{equation}
L^{acc}_{disc} = \frac{3G M_h dM/dt}{2} \left( \frac{1}{r_1} [1-\frac{2}{3}(\frac{r_{\star}}{r_1})^{1/2}]-  \frac{1}{r_2}[1-\frac{2}{3}(\frac{r_{\star}}{r_2})^{1/2}] \right)
\end{equation}
\noindent
where $r_{\star}$,  $r_1$ and $r_2$ are the primary, inner-disc and outer-disc radius, respectively, and $G$  the gravitational constant \citep{2002apa..book.....F}. For the parameters of HD\,170582 ($r_1$ =  $r_{\star}$ and $r_2$ = 3.8 $r_{\star}$ from M15) we obtain $L^{acc}_{disc}$ = 19.6 $L_{\odot}$.


The disc luminosity can also be calculated using the M15 light-curve model based on the Nelder-Mead simplex algorithm (see e.g. Press et al. 1992) with optimizations described by Dennis and Torczon (1991), and the model of a binary system with a disc. For more detail see e.g. Djura\v{s}evi\'{c} (1992). We call this figure the {\it observed} disc luminosity; it can be determined with typical accuracy less than 10\%.
For  HD\,170582 we obtain 1564 $L_{\odot}$, much higher than the accretion luminosity  calculated with Eq.\,4. 

The above result moved us to compare  the accretion luminosity of all well-studied DPVs with their disc luminosities, the last ones inferred from the light curve models that we have found in recent years. 
These data are shown in Table\,4 and illustrated in Fig.\,11. 
We find that for the 5 studied DPVs the observed disc luminosity is between 14 and 54\% of the primary luminosity, and always much larger than the accretion luminosity.  
In addition, the observed disc luminosity scales with the primary mass whereas the accretion luminosity does not.

We conclude that the discs for the 5 well-studied DPVs are not accretion powered. We notice that this result rests on the value for $dM/dt$, that was determined through comparison of the observed system parameters with
grids of synthetic evolutionary paths. We also notice that for having similar accretion and disc luminosities the mass transfer rate  would increase by several orders of magnitude, reaching unrealistic values which should produce large but undetected orbital period changes (Table\,4, e.g. Eggleton 2011). 
For the above reason, we are confident of the difference between  observed and accretion disc luminosity, and consequently, one should not use the accretion disc theory  to derive mass transfer rates for these systems.

\section{Discussion}


Doppler tomograms for DPVs have been previously published for AU\,Mon (Atwood-Stone et al. 2012, Richards et al. 2014), and V\,393 Sco (Mennickent et al. 2012a, 2012b).
For AU\,Mon the disc is clearly visible in the Doppler tomograms in all quadrants;  Atwood-Stone et al. (2012) identify ``an asymmetric component of an elliptical accretion disk, and material moving at sub-Keplerian velocities that provides excess H$\alpha$ emission, plausibly associated with the continuation of the mass transfer stream beyond the splash site'' and this result is confirmed by Richards et al. (2014). The H$\alpha$ Doppler map of  V\,393\,Sco (at high state) reveals a ring-like emissivity distribution  with enhanced 
emission in the 1st and 4th quadrants. 

The case of the DPV HD\,170582 is considered in this paper; for the first time Doppler maps of a DPV are separately shown for the high and low stage.  The maps reveal a horseshoe pattern with increased emission at the 1st and 4th quadrants. 
While emissivity at the 1st quadrant can be associated to the region of penetration of the stream into the disc, emission at the 4th quadrant can be interpreted as a shock generated by the disc material encountering the stream after  encircling  the primary, as hydrodynamical simulations predict \citep{2001PThPh.106..729F}.  The position of the first 
structure  coincides with  one of the bright spots found after the light curve analysis (M15).

As mentioned above very few DPVs have Doppler tomographic reconstruction, therefore we still cannot extract significant conclusions of line emissivity distributions for DPVs as a class. However, the three cases discussed  so far show the  variety of patterns that can be found, from whole emitting discs (AU\,Mon) to emission located at specific quadrants (HD\,170582).
It is interesting that in the case of  the eclipsing systems AU\,Mon \citep{2014BarriaPhD}, V\,393\,Sco \citep{2012arXiv1205.6848M} and DQ\,Vel \citep{2013A&A...552A..63B} larger Balmer emission is found at high state.  In HD\,170582, seen at intermediate inclination,   the H$\alpha$ emission at the stream-disc impact region during high state decreases, compared with the H$\beta$ emission, something that could be interpreted as evidence of optically thicker material produced by a larger mass transfer rate. If this tendency is confirmed in other DPVs, it could be an indicator that the long cycles are modulated by a variable mass transfer rate. In this context  it is worth noticing that the larger Balmer emission observed at the high state of V\,393 Sco has been interpreted as the result of  a bipolar wind emerging from the stream-disc impact region \citep{2012arXiv1205.6848M}.


 DPV discs seems to be stable, optically thick and relatively small, with radial extension a few times the stellar radius  \citep{2015arXiv151005628M}. 
Since DPVs are tangential impact systems, these discs are probably formed by saturation of the primary after rapid spin-up till critical velocity \citep{2015arXiv151005628M}. To our knowledge,  no previous study has been done
to inquire about the  source of luminosity  for these discs. We have found that they are not accretion powered, like discs around compact objects. This is reasonable due to the smoother gravitational potential well. However, they are relatively luminous, 
as much as 50\% the primary luminosity.  One possible heating source for the disc is the primary star, through capture and reprocessing of their high-energy photons, others possible sources are the shock produced by the gas stream and the disc dynamics.

\section{Conclusions}

Most of the spectra of HD\,170582 used in this paper were obtained with resolving power 40 000 between years 2008-2013 and were presented for the first time by M15. 
In this work we present a complementary analysis, studying the orbital and long-term behavior of H$\alpha$ and H$\beta$ emission lines obtained after disentangling the donor spectrum. We also explore the system evolutionary stage.  Critical in
our methodology is the assumption that the line emissivity can be recovered by subtracting the donor absorption spectrum at every orbital phase. In addition, comparison 
of the system parameters with the grid of binary star evolutionary tracks is limited by the ad-hoc treatment of mass loss by the  models. This point might be not an important issue, since the best model is a conservative one.
Keeping in mind these limitations, we have arrived  at the following conclusions: \\

\begin{itemize}
\item The line emission velocities are not associated  with the stable and optically thick disc contributing to the orbital light variability, but to outer optically thinner circumstellar regions. 
\item Two bright regions, one in the 1st and  another in the 4th quadrant, are visible in the Doppler tomograms. They can be interpreted as shock regions arising from the gas dynamics in the system, as indicated by published hydrodynamical simulations of semidetached close binaries; the penetration of the gas stream into the disc and the hitting of disc material back into the stream. 
\item  We find a change in the emissivity properties of the material located in the stream-disc impact region: at high stage the H$\alpha$ emission is weaker than in low stage compared with H$\beta$ emission. This fact might indicate optically thicker material at high state, possibly caused by a larger mass transfer rate. If this tendency is confirmed in other DPVs, it could be an indicator that the long cycles are modulated by variable mass transfer rates.
\item The disc luminosity, inferred from the light curve model, is larger  than expected due to accretion. Therefore, other heating sources are likely acting in the disc material, i.e. reprocessing of the primary high-energy photons or heating by shock regions.
\item The disc luminosity for 5 well-studied DPVs scales with the primary mass and is always much larger than the luminosity inferred from the accretion disc theory and the published mass transfer rates.
\item For the above reason, we should not use the assumption of accretion driven disc luminosity to infer values for the mass transfer rate in DPVs.   
\item  We find the best model for the system among a grid of evolutionary models for Algol-type binaries. It is found in a Case-B conservative mass transfer stage, with an age of 7.68 $\times$ 10$^{7}$ yr and a mass transfer rate of 1.6
 $\times$ 10$^{-6}$ M$_{\odot}$ yr$^{-1}$.
\end{itemize}

\begin{table*}
\centering
 \caption{For DPVs we give the accretion luminosity according to Eq.\,4 and the disc luminosity extracted from the light curve models. 
 Primary mass, radius and luminosity and disc radius and luminosity 
 are given in solar units. 
 {\rm $dM_c/dt$} is the (probably unrealistic) mass transfer rate needed to match observed and accretion disc luminosity.
 Except last four columns, data are from M15 and reference therein.} 
 \begin{tabular}{@{}lrccccrrcrrrc@{}} \\
 \hline
 \hline
{\rm DPV}  &{\rm $M_h$} &  {\rm $eM_h$} &{\rm $R_h$ } & {\rm $eR_h$} & {\rm $dM/dt$ ($M_{\odot} yr^{-1}$)} &{\rm $L_h$}& {\rm $R_d$}&   {\rm $eR_d$} &{\rm $L^{acc}_d$} &{\rm $L_d$} &{\rm $L_d/L_h$} &{\rm $dM_c/dt$ ($M_{\odot} yr^{-1}$)} \\
 \hline
OGLE\,05155332-6925581 &	9.1	&0.5&	5.6	&0.2	&3.1E-6	&10471&	14.1	&0.5	&24.45	&4225 &0.40&5.0E-4\\
HD\,170582	&9	&0.2	&5.5	&0.2	&1.6E-6	 &2884	&20.8&	0.3	&	19.65&	1564&0.54&2.0E-4\\
V393\,Sco	&7.8	&0.5	&4.1	&0.2&	9.5E-9&	1148 &	9.7	&0.3	&0.07&	460&0.40&1.0E-4\\
DQ\,Vel	&7.3	&0.3	&3.6	&0.2&	9.8E-9	&1380&	12.9	&0.3	&	0.06&	413&0.30&1.0E-4\\
AU\,Mon	&7	&0.3&	5.1	&0.5&	7.6E-6	&1380&	12.7&	0.6		&49.77&	197&0.14&3.0E-5\\
\hline
\end{tabular}
\end{table*}

\begin{figure*}
\scalebox{1}[1]{\includegraphics[angle=0,  viewport = 105 210 540 750,width=14cm]{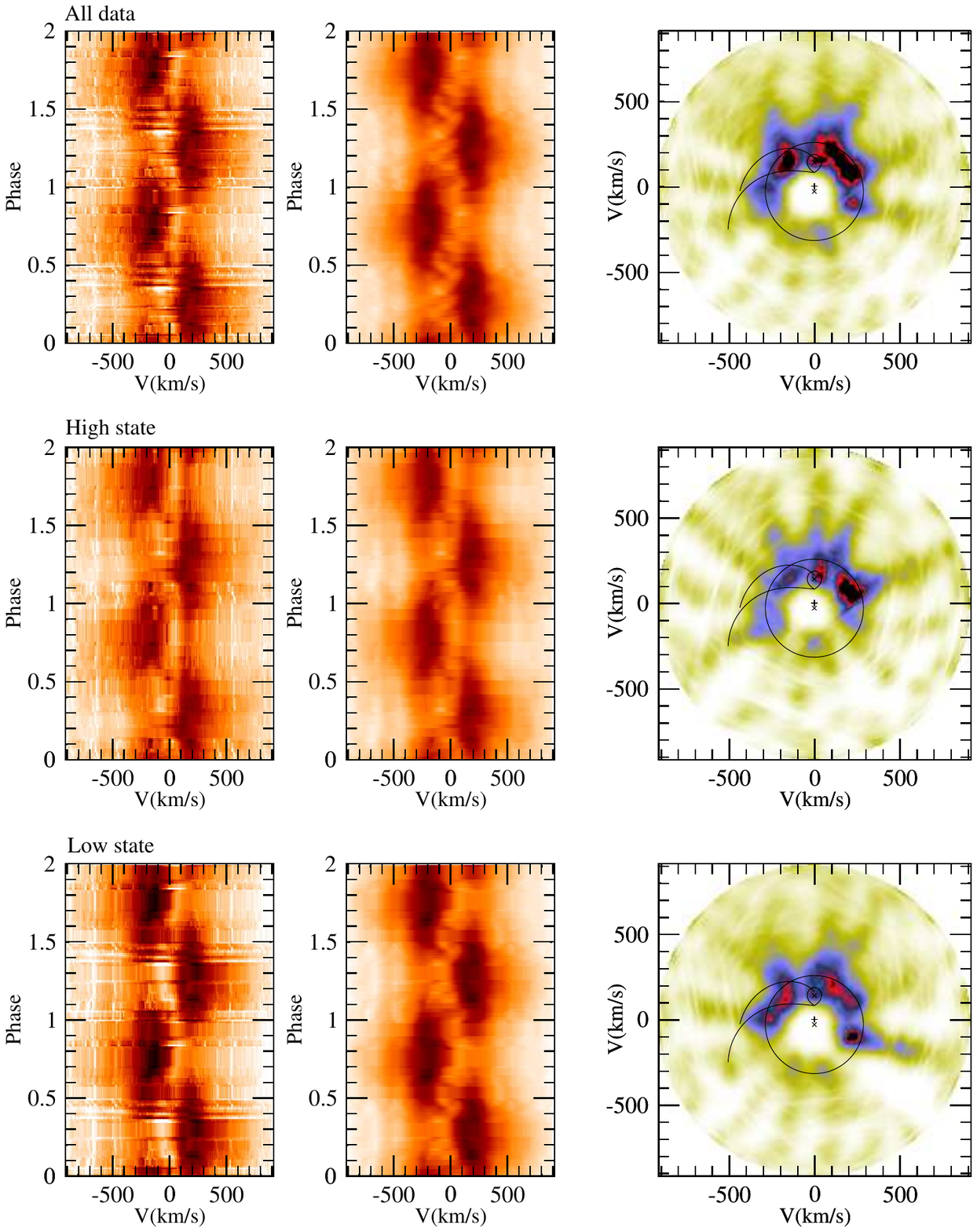}}
\caption{Phased time series observed and reconstructed spectra around H$\alpha$ line folded with the orbital period of the system and corresponding Doppler maps. 
Most spectra at the high (low) stage are in 0.7 $<$ $\Phi_l$ $<$ 1.0 (0.4 $<$ $\Phi_l$ $<$ 0.6).
The orbital period of $P_{orb}$ = 16.87 days, the primary mass of $M_1$ = 9.0 \msun, inclination angle 67\fdg4 and the mass ratio of $q$ = 0.21 from M15  are used to overlay positions of the stellar components on the Doppler maps. $\Phi$ = 0.0  corresponds to the inferior conjunction of the donor. 
The loci for the center of mass for both stellar components, the theoretical ballistic gas stream and the Keplerian velocity at the stream are marked on the tomograms. The circle represents the Keplerian velocity of the disc outer radial edge as inferred from the light curve model (M15).
 The filling of missing phases in the folded spectra is used for best presentation.
}
  \label{x}
\end{figure*}

\begin{figure*}
\scalebox{1}[1]{\includegraphics[angle=0, viewport = 105 210 540 750, width=14cm]{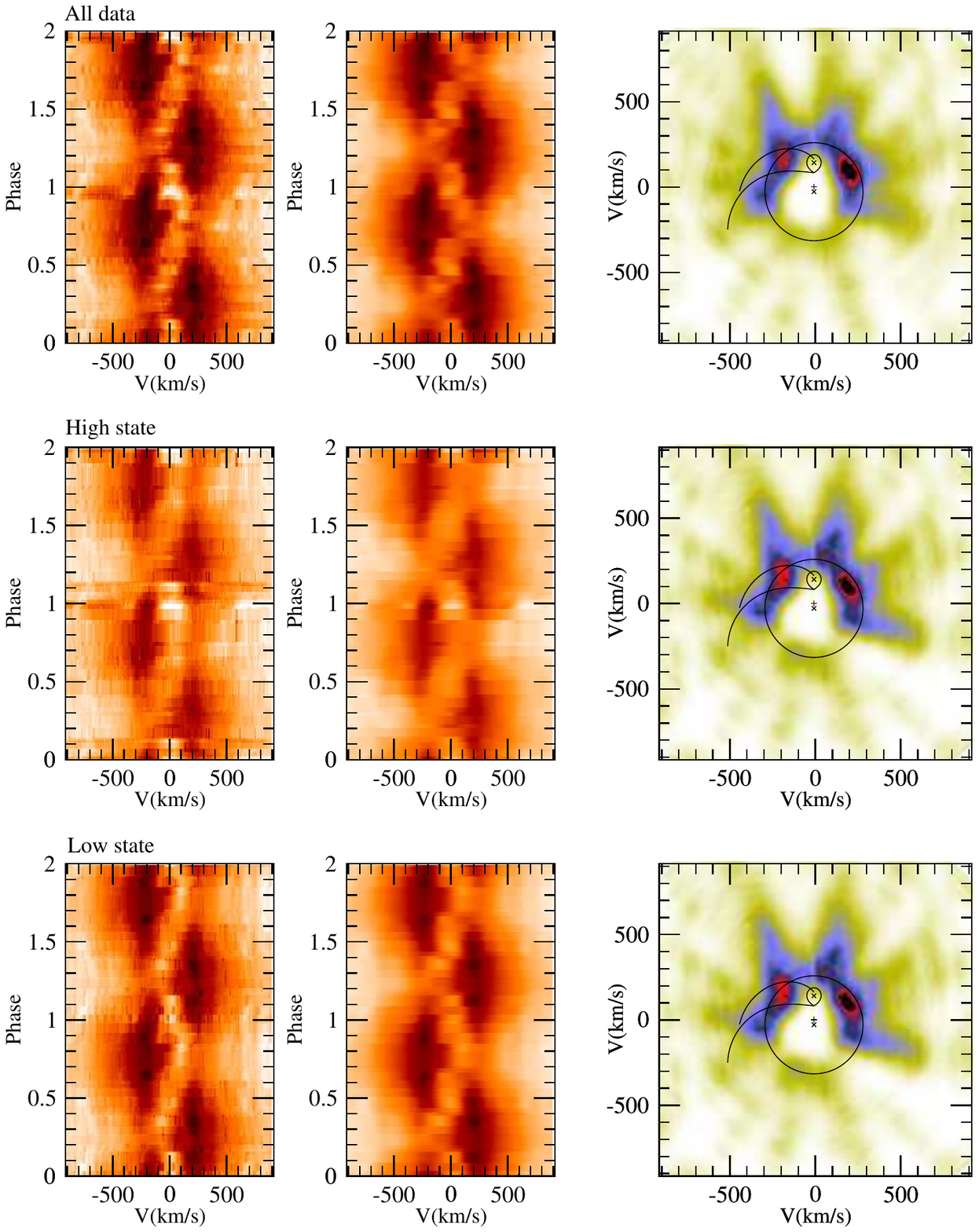}}
\caption{Phased time series observed and reconstructed spectra around H$\beta$ line folded with the orbital period of the system and corresponding Doppler maps. 
Most spectra at the high (low) stage are in 0.7 $<$ $\Phi_l$ $<$ 1.0 (0.4 $<$ $\Phi_l$ $<$ 0.6).
The orbital period of $P_{orb}$ = 16.87 days, the primary mass of $M_1$ = 9.0 \msun, inclination angle 67\fdg4 and the mass ratio of $q$ = 0.21 from M15  are used to overlay positions of the stellar components on the Doppler maps. $\Phi$ = 0.0  corresponds to the inferior conjunction of the donor. 
The loci for the center of mass for both stellar components, the theoretical ballistic gas stream and the Keplerian velocity at the stream are marked on the tomograms. The circle represents the Keplerian velocity of the disc outer radial edge as inferred from the light curve model (M15).
 The filling of missing phases in the folded spectra is used for best presentation.
}
  \label{x}
\end{figure*}


\begin{figure}
\scalebox{1}[1]{\includegraphics[angle=0,width=8cm]{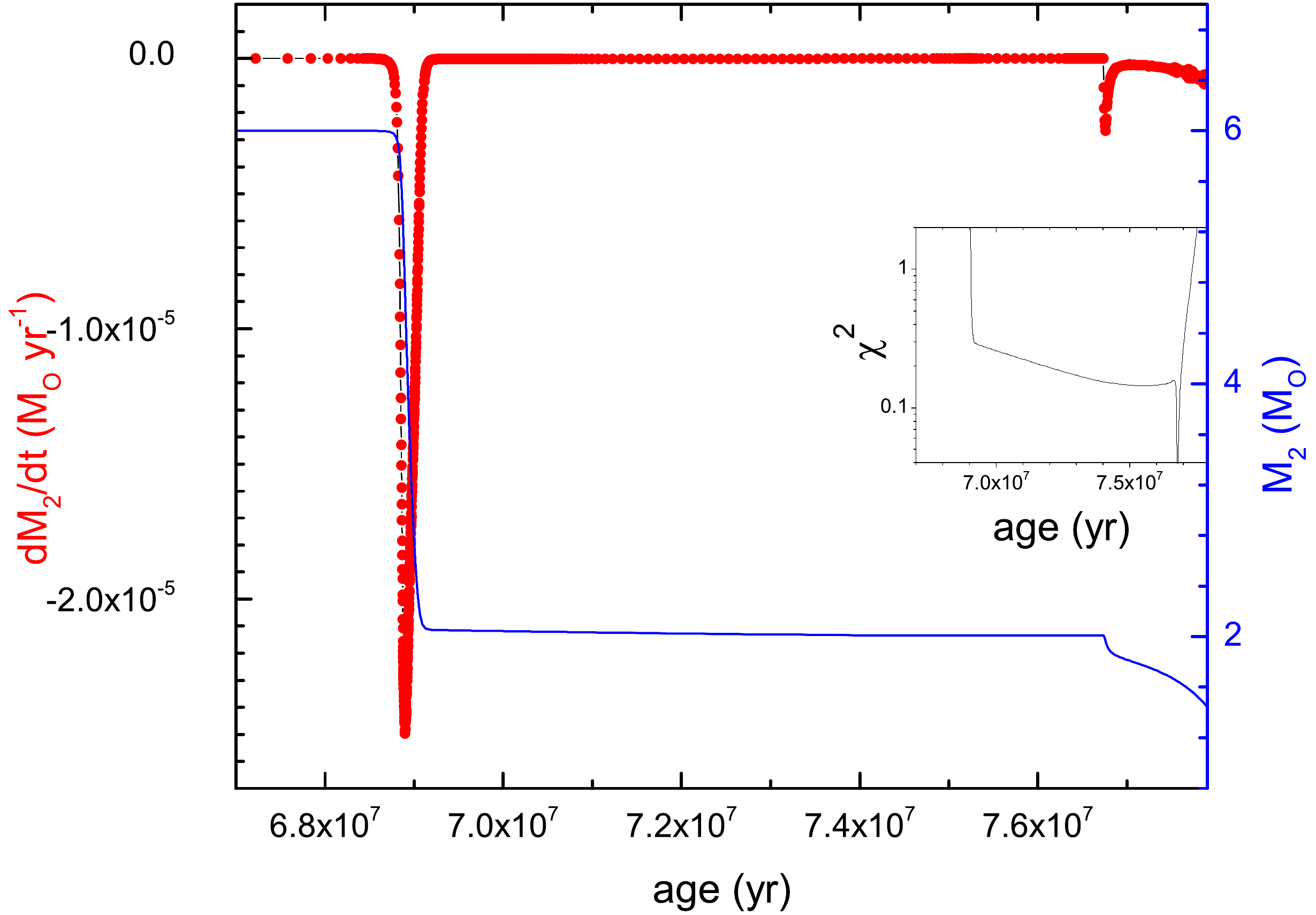}}
\scalebox{1}[1]{\includegraphics[angle=0,width=8cm]{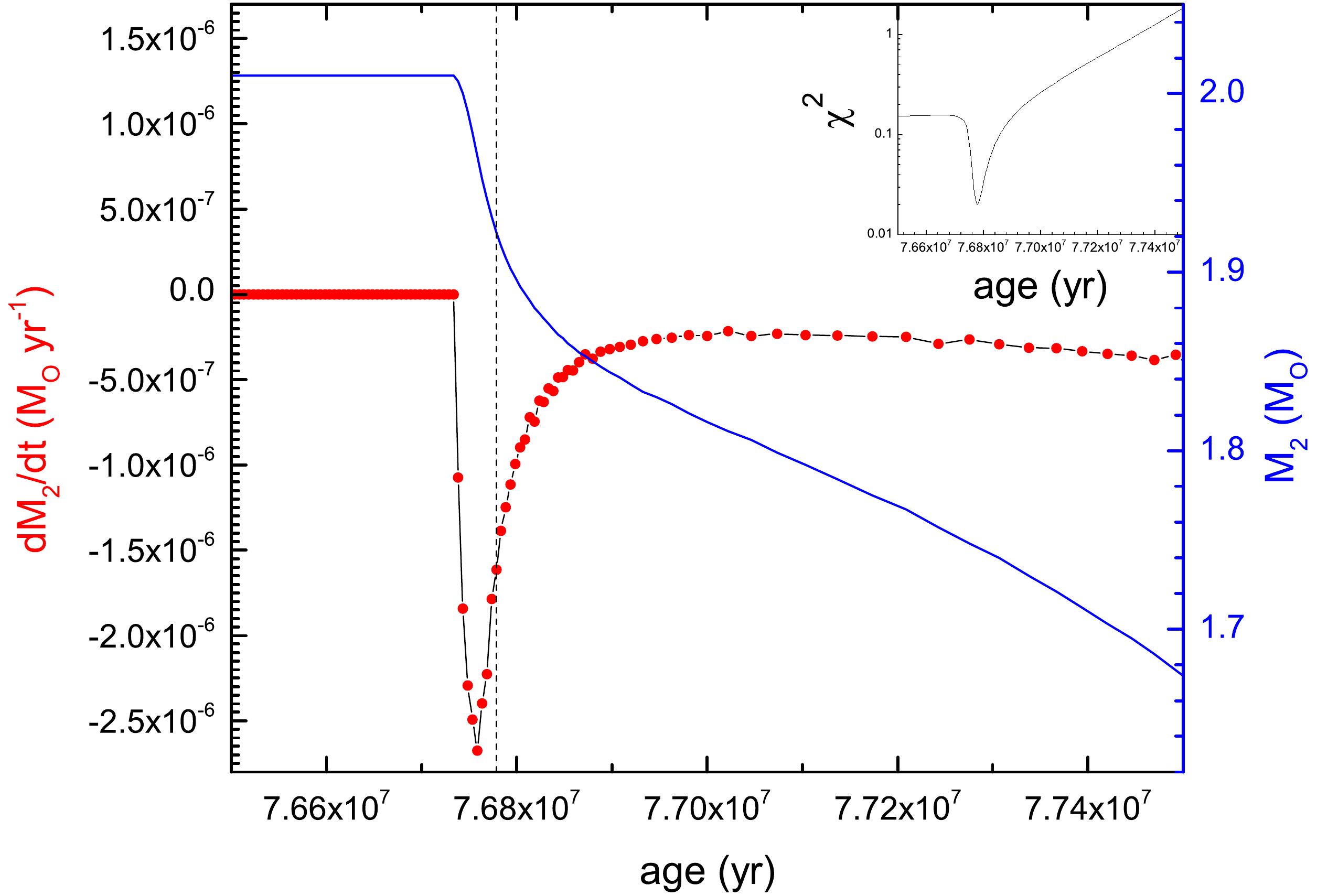}}
\caption{The mass transfer rate and donor mass as a function of time for the best model. The  lower panel  is an enlarged version of the upper graph, indicating the current epoch with a vertical dashed line. This epoch corresponds to the minimum of the $\chi^{2}$ represented in the inset graphs. }
  \label{x}
\end{figure}

\begin{figure}
\scalebox{1}[1]{\includegraphics[angle=0,width=8cm]{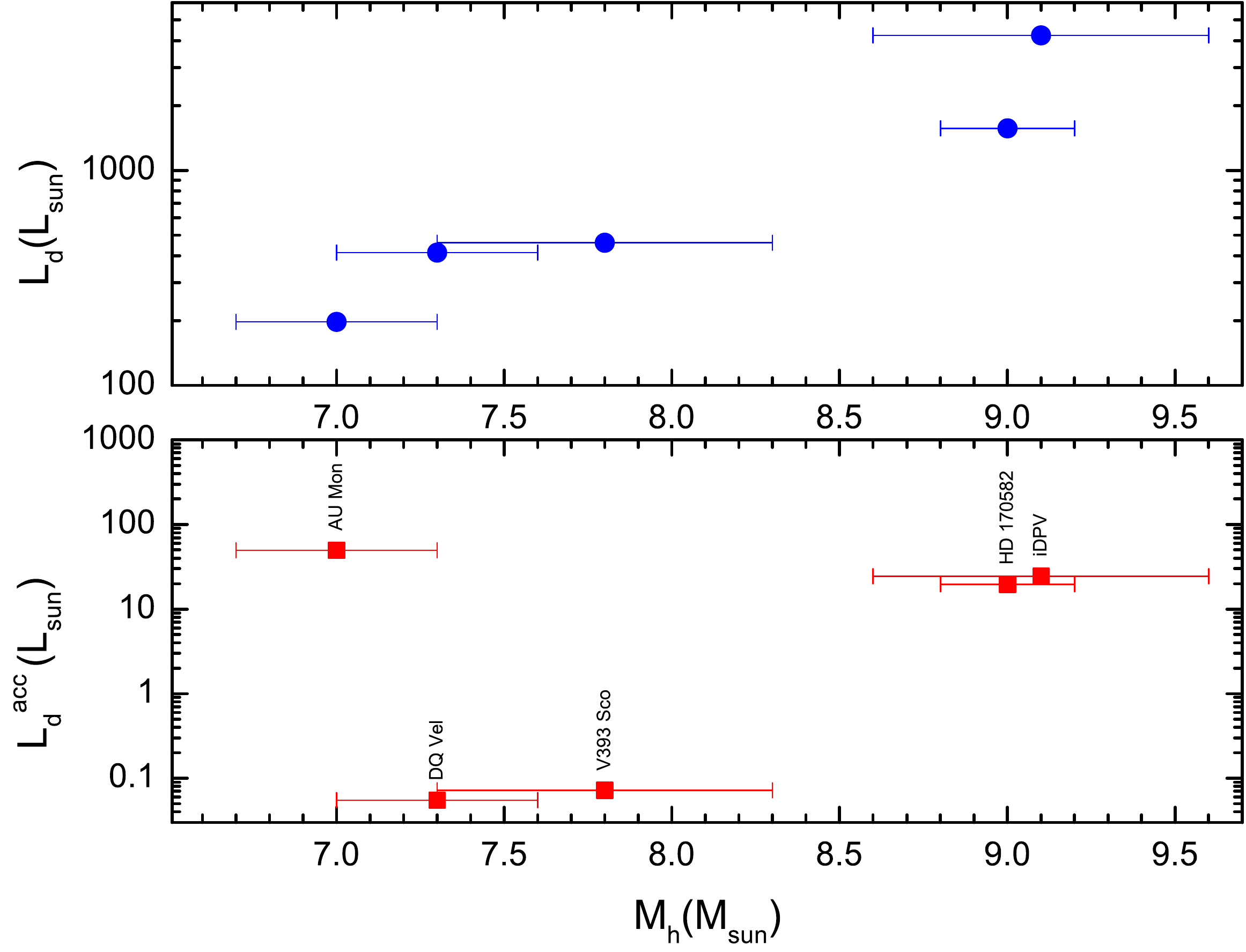}}
\caption{The bottom panel shows the accretion luminosity for DPVs, as derived from the mass transfer rate of the best binary model for the present  system and stellar parameters. 
The upper panel shows the disc luminosity inferred from the  light curve analysis. iDPV stands for OGLE05155332-6925581 and data are from Table\,4.}
  \label{x}
\end{figure}

\section*{Acknowledgments}


 We acknowledge an anonymous referee for useful comments regarding the first version of this paper.
R.E.M. acknowledges support by VRID-Enlace 214.016.001-1.0 and the BASAL Centro de Astrof{\'{i}}sica y Tecnolog{\'{i}}as Afines (CATA) PFB--06/2007. 
S. Z. acknowledges support from DGAPA/PAPIIT project IN100614 and CONACyT grant CAR 208512.
G. D. gratefully acknowledges the financial support of
the Ministry of Education and Science of the Republic of Serbia
through the project 176004, Stellar physics.






\bsp	
\label{lastpage}
\end{document}